\documentclass[letterpaper, 10 pt, conference]{ieeeconf}
\IEEEoverridecommandlockouts
\usepackage{amsmath,amssymb,amsfonts}
\usepackage{algorithmic}
\usepackage{graphicx}
\usepackage{textcomp}
\usepackage{booktabs}
\usepackage{wrapfig}
\usepackage[table]{xcolor}

\usepackage[most]{tcolorbox}
\tcbuselibrary{breakable}

\usepackage{listings}

\definecolor{codegreen}{rgb}{0,0.6,0}
\definecolor{codegray}{rgb}{0.5,0.5,0.5}
\definecolor{codepurple}{rgb}{0.58,0,0.82}
\definecolor{backcolour}{rgb}{0.96,0.96,0.96}
\definecolor{bestcol}{RGB}{255,127,80} 
\definecolor{secondcol}{RGB}{255,215,100}

\lstdefinestyle{emnlpstyle}{
    backgroundcolor=\color{backcolour},   
    commentstyle=\color{codegreen}\itshape,
    keywordstyle=\color{blue}\bfseries,
    numberstyle=\tiny\color{codegray},
    stringstyle=\color{codepurple},
    basicstyle=\ttfamily\scriptsize, 
    breakatwhitespace=false,         
    breaklines=true,                
    captionpos=b,                  
    keepspaces=true,                 
    showspaces=false,                
    showstringspaces=false,
    showtabs=false,                  
    tabsize=2,
    frame=single,                  
    rulecolor=\color{black!30}      
}

\def\BibTeX{{\rm B\kern-.05em{\sc i\kern-.025em b}\kern-.08em
    T\kern-.1667em\lower.7ex\hbox{E}\kern-.125emX}}
\begin{document}

\title{ROSTrans: Empirically Benchmarking LLMs for ROS1-to-ROS2 Migration}

\author{
\authorblockN{
\parbox{\textwidth}{\centering\normalsize
Yichen Zhao\authorrefmark{1},
Minghao Shao\authorrefmark{2}\authorrefmark{3},
Zhuoyan Yu\authorrefmark{1},
Yijun Shen\authorrefmark{1},
Boyuan Chen\authorrefmark{2}\authorrefmark{3},
Yik-Cheung Tam\authorrefmark{1},
Muhammad Shafique\authorrefmark{3}
}
}
\authorblockA{
\parbox{0.95\textwidth}{\centering\normalsize
\authorrefmark{1}Center for Data Science, NYU Shanghai, China\hspace{1.5em}
\authorrefmark{2}NYU Tandon, USA\hspace{1.5em}
\authorrefmark{3}NYU Abu Dhabi, UAE\\
\{yz10469, shao.minghao, ys6126, zy3077, bc3194, wilson.tam, muhammad.shafique\}@nyu.edu
}
}
}

\maketitle

\begin{abstract}
Although large language models have shown strong capabilities in
domain-specific code generation, their performance in robotics software development, especially for the Robot Operating System (ROS), remains largely unexplored. ROS code generation presents distinctive challenges, including complex inter-node coordination, middleware configuration, and lifecycle semantics that go far beyond standard coding benchmarks, and many important correctness properties become observable only once migrated nodes are launched and communicate. We present ROSTrans, a benchmark for evaluating LLMs on
ROS1-to-ROS2 code migration, a representative and practically important setting for ROS code generation. ROSTrans comprises 108 scaffolded completion tasks drawn from a diverse set of 70+ open-source ROS1 repositories and organized into a hierarchical
taxonomy assessing interface-level, behavioral-level and system-level
correctness. For every task we establish an executable reference: a verified ROS2 implementation, a runtime test suite that launches the nodes and exercises their topics, services, actions and parameters, and a container ensuring reproducibility. Candidates are scored by injecting only their target files into an otherwise fixed copy of the reference package, so a candidate can neither supply nor weaken the tests it is judged by. Evaluating six frontier LLMs over five stochastic generations per task, we find pass@1 below 10.9\%, and that 56.5\% of all samples build successfully and then fail at runtime,and 32.0\% fail runtime assertions on the migrated interfaces, a class of error no compile-only oracle can detect. Execution-guided repair further improves task success, reaching 49.1$\%$ and 47.2$\%$ for Claude Opus 4.6 and GPT-5.5 after three rounds, respectively.

\end{abstract}

\section{Introduction}
\label{sec:introduction}

Large language models have evolved into general-purpose reasoning engines, showing strong performance in content understanding and knowledge-grounded question answering~\cite{shao2024survey}. Beyond natural language processing, LLMs are now widely applied to code generation across specialized domains, from hardware design and verification~\cite{wang2024llms,blocklove2023chip,fu2026synthesis} to quantum circuit generation~\cite{basit2025pennycoder}. Models such as Codex, CodeLlama, and DeepSeek-Coder achieve strong results on mainstream programming languages~\cite{chen2021evaluating}, while recent work extends code intelligence to repository-scale understanding, cross-file patch generation~\cite{jimenez2024swe}, and vulnerability detection and repair~\cite{yu2025patchagent,xi2025trace}. LLM-based code generation is therefore advancing along two closely related directions, broadening domain coverage while deepening software engineering reasoning from isolated functions to complete project lifecycles.

Despite this progress, LLM capabilities in robotics software development remain insufficiently explored. The ongoing transition from the legacy ROS1 ecosystem to ROS2 provides a representative and practically important setting for ROS code generation, yet it requires more than API renaming. It demands rethinking communication middleware, node lifecycle management, execution semantics, and system organization~\cite{macenski2022robot}. Robotics code must also satisfy complex inter-node coordination, real-time Quality of Service constraints, and execution semantics that go beyond the scope of standard coding benchmarks~\cite{stavrinos2021ros2}. These challenges are further amplified by the heterogeneity of ROS packages, distributions, build systems, and application domains, where high-quality public code is limited and fragmented. As a result, data-driven models struggle to acquire the structural priors needed for reliable robotics programming~\cite{guiochet2017safety}. Yet the community still lacks standardized benchmarks for evaluating LLM performance on ROS1-to-ROS2 migration.

\begin{figure}[t]
    \centering
    \includegraphics[width=\linewidth]{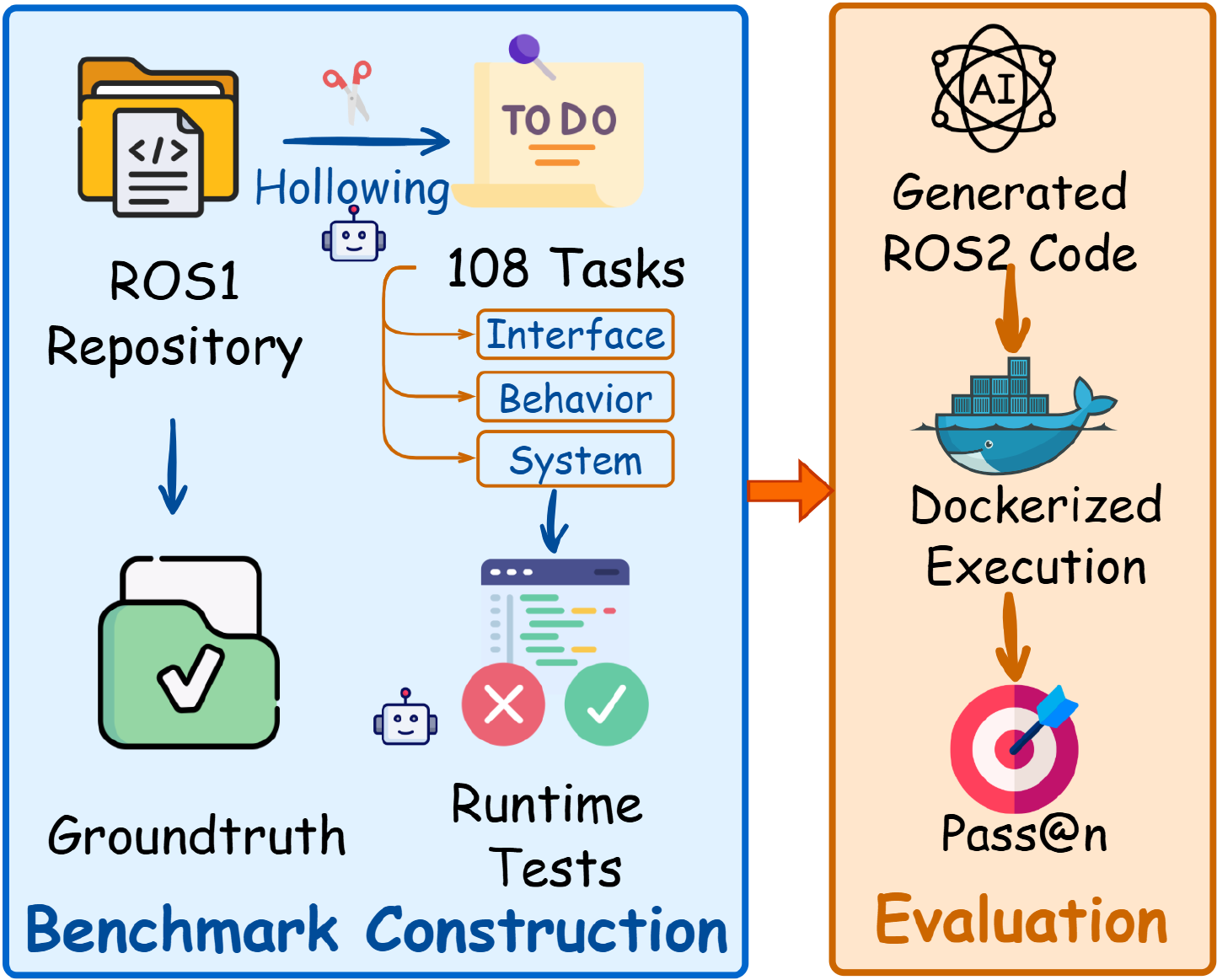}
    \caption{ROSTrans overview. ROS1 repositories are transformed into 108  tasks via LLM-assisted hollowing, then evaluated via pass rate.}
    \label{fig:overview_figure}
\end{figure}

To bridge this gap, we present ROSTrans\footnote{Benchmark source code and evaluation framework are open source at: github https://github.com/NYU-Serendipity-LLM4Coding/ROSTrans}, a comprehensive benchmark evaluating LLMs on robotics-specific code migration. 
Given a partially
specified ROS1 codebase and a task description, the goal is to generate
functionally equivalent ROS2 code with correct inter-node communication and
logical execution flow. ROSTrans derives 108 tasks from 70+ open-source ROS1
repositories, organized by a hierarchical taxonomy measuring interface, behavioral and system-level correctness. All tasks use
scaffolded code completion to isolate architectural reasoning from boilerplate
generation. For each task we establish an executable reference by translating
the hollowed sources, building the result in a container, running the tests it
must satisfy, and feeding failures back for repair until the package builds and
passes. Candidates are then scored by a hybrid-injection protocol in which only
the candidate's target files replace their counterparts in an otherwise fixed
copy of the reference package, leaving the tests and the build configuration
outside the candidate's reach. Our contributions are as follows.
 
\begin{itemize}
\item We propose ROSTrans, a benchmark dedicated to
ROS1-to-ROS2 code migration in which every task is decided by building and
running the migrated package, across three levels of architectural scope.
\item We describe a construction and evaluation protocol that makes such a
benchmark auditable: references are established by execution rather than
asserted, test adequacy is checked by mutation, and the evaluation path itself
is validated by scoring the reference as a candidate and by injecting
deliberately invalid content.
\item We evaluate six frontier LLMs and find uniformly low success rates, with
the majority of failures occurring after a successful build, and characterize
where and why the migrations break down.
\end{itemize}

\section{Background}
\label{sec:background}
The transition from ROS1 to ROS2 represents a fundamental architectural shift in robotic middleware. While ROS1 relied on a centralized master node and custom TCPROS protocols, ROS2 adopts the industry-standard Data Distribution Service as its communication backbone, enabling decentralized discovery and real-time capabilities~\cite{macenski2022robot}. This redesign introduces Quality of Service profiles, explicit node composition, and automatic latency management~\cite{stavrinos2021ros2, zhang2024comparison, blass2021automatic, macenski2023impact}, yet it also means that legacy migration must go far beyond API renaming to encompass wholesale restructuring of execution semantics and network topologies.

Large Language Models have demonstrated strong capabilities in generalized code synthesis, yet their performance degrades sharply in low-resource or highly specialized domains~\cite{gu2025effectiveness, joel2025survey}, especially when the target framework requires deep middleware structural priors and physical constraints~\cite{legacy2024, dong2025towards}. Moreover, program translation across architectural boundaries demands strict semantic and behavioral consistency rather than superficial lexical mapping~\cite{du2023beyond}. Without such structural context, code generation approaches frequently produce output that is syntactically correct but functionally invalid in robotic settings, underscoring the need for specialized benchmarks.

Automated code evaluation typically relies on execution-based oracles, in which generated code is executed against test suites and judged by functional correctness~\cite{chen2021evaluating, jimenez2024swe}. For robotic middleware, however, executing isolated functions is insufficient, since node discovery, message delivery under Quality of Service settings, and callback scheduling only manifest when nodes are launched and communicate. ROS therefore supports integration-level runtime testing, in which the nodes under test are launched, inputs are published, and assertions are made on the observed messages, services, or parameters, as provided by rostest in ROS1 and launch\_testing in ROS2~\cite{rostest, ros2launchtesting}. Such tests remain difficult to design and automate in practice~\cite{afzal2020study}, and their validity depends on test adequacy, since tests that do not exercise the intended behavior can admit incorrect solutions~\cite{liu2023your}. Mutation testing provides a standard means of assessing adequacy by injecting faults into the program under test and checking whether the tests detect them~\cite{demillo1978hints, jia2011analysis}.

Runtime testing of robotic software further depends on its execution environment. Unlike conventional software where failure carries negligible physical risk, robotic systems operate in embodied environments where software errors can cause hardware damage~\cite{guiochet2017safety}, making reliable validation before deployment essential. However, a ROS runtime test requires a precisely configured stack of distribution, package dependencies, build tools, and middleware implementation, and results obtained on one host may not be reproducible on another. Containerization addresses this dependency by packaging the complete software stack into an isolated and reproducible image~\cite{boettiger2015introduction}, and is commonly used for developing and deploying ROS systems~\cite{white2017ros}. These properties make containerized runtime testing a natural basis for evaluating migrated ROS code at scale.

\section{Benchmark Construction}
\label{sec:benchmark}

\begin{figure*}[t]
    \centering
    \includegraphics[width=\textwidth]{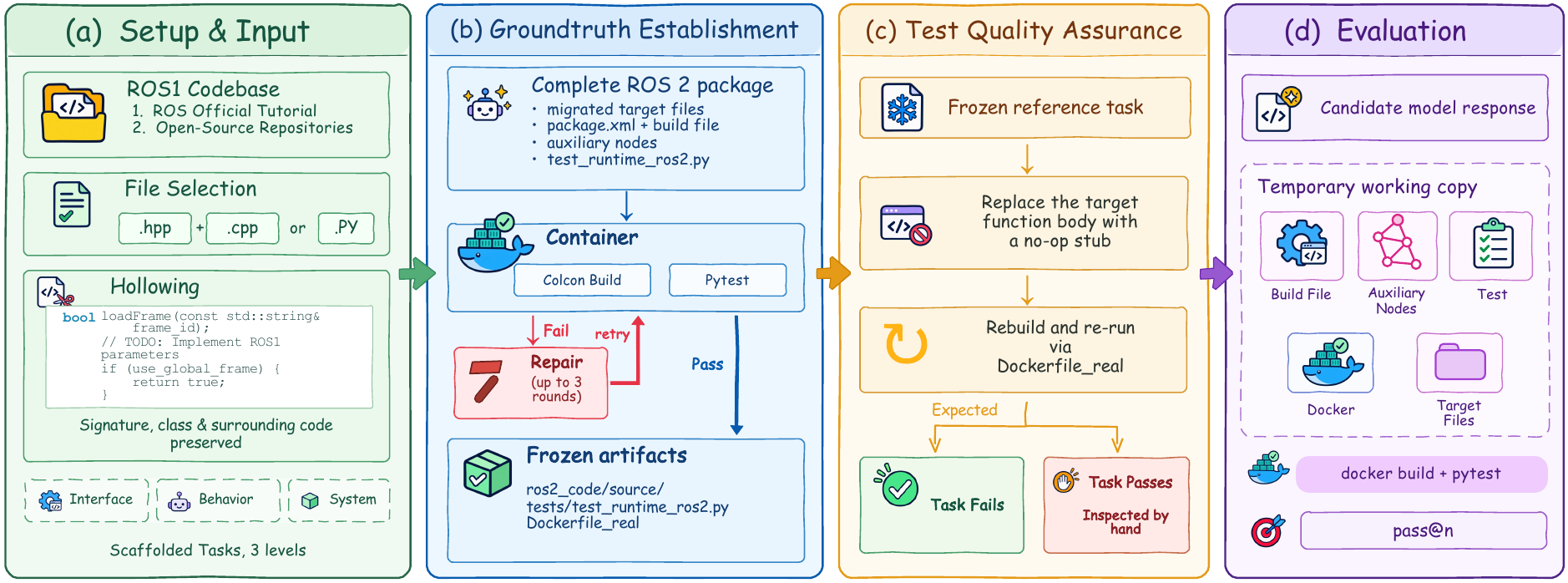}
    \caption{\textbf{Pipeline of ROSTrans}. (a) Benchmark Construction: coherent file groups are selected from legacy ROS~1 code and hollowed into scaffolded tasks with structured TODO markers across three capability levels. (b) Reference Establishment: each task is translated to ROS~2 and verified by building and running it inside a container, with build and test logs fed back for repair until the task builds and its runtime tests pass. (c) Test Quality Assurance: the hollowed function is replaced by a no-op stub and the task is re-run to check that its tests detect the change. (d) Candidate Evaluation: only the candidate's target files are injected into an otherwise fixed copy of the reference package, yielding a binary outcome and the stage at which failure occurred.}
    \label{fig:my_wide_image}
\end{figure*}

\subsection{Sources and Selection}

We start from legacy ROS~1 codebases and migration resources, hollow the relevant code, add runtime tests, and verify each task by execution. Task objectives require inferring global topology from local contexts, including coordinated node behavior, distributed callback execution, parameter handling, and multi-node message exchange. Complete repositories are converted by masking key functions or logic blocks. Tutorial sources are first turned into compact ROS~1 nodes and then scaffolded. When the target behavior depends on wider system interactions, we add auxiliary nodes or external tests to ground the task in the intended ROS behavior.

Each task is built from a file group that forms a coherent unit of robotic behavior and can be exercised on its own. The upstream source of each task is documented in its accompanying README and will be released with the code.

\subsection{Capability Levels}

ROSTrans measures ROS coding capability along three levels of semantic scope. \textbf{Interface-level} tasks test basic communication and configuration patterns, including action servers, parameter servers, publisher-subscriber communication, and service-client interaction. \textbf{Behavior-level} tasks require coordinating multiple ROS components within one program, such as perception-control pipelines, sensor processing and state machines. \textbf{System-level} tasks test architectural reasoning across files, nodes, and distributed ROS semantics, covering multi-node coordination and simulation under strict global constraints.

\begin{figure}[htbp]
  \centering
  \includegraphics[width=\linewidth]{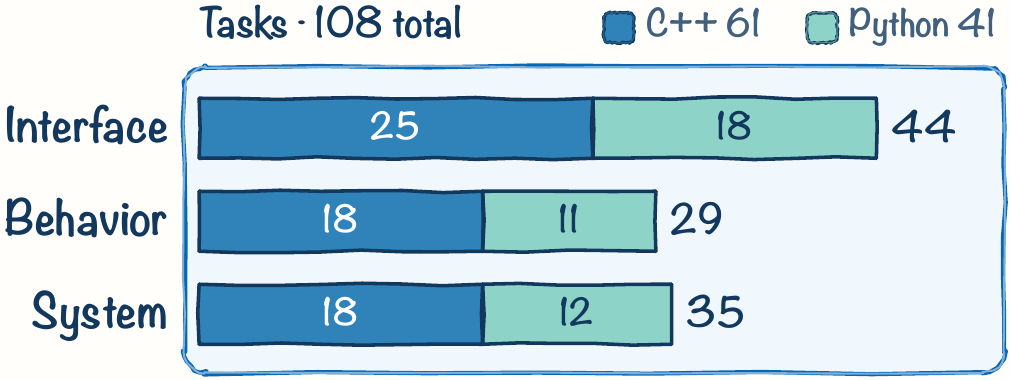}
  \caption{Task statistics by capability level. Language categorization is based on target file extensions (mixed tasks are counted as C++). Modifications and hollowed lines are measured via TODO markers in the scaffolded source.}
  \label{fig:tasks}
\end{figure}

ROSTrans contains 108 ROS coding tasks organized by capability level: 44 interface-level, 29 behavior-level, and 35 system-level tasks. Together, they cover localized ROS API usage, component-level behavior, and higher-level reasoning for multi-node coordination, navigation, manipulation, and simulation integration. The benchmark spans diverse ROS software, including 61 C/C++ tasks, 41 Python tasks, and 6 declarative configuration tasks covering launch, xacro, YAML, and robot descriptions. Task difficulty ranges from short parameter or interface updates to multi-hundred-line completions. Figure~\ref{fig:tasks} summarizes the benchmark.

\subsection{Task Scaffolding}

The scaffold isolates architectural reasoning from noisy boilerplate generation, keeping surrounding context while removing key implementation to recover intended ROS~2 behavior. Function signatures, enclosing class and namespace structure, imports and includes, and all non-target code are preserved; only the body of the targeted region is removed. In its place we insert a block marker, opened by a \texttt{TODO} comment and closed by an \texttt{END OF TODO} comment, that states in natural language what the region must accomplish after migration. Markers may also carry explicit style constraints, for instance requiring a particular ROS~2 API form where several are equivalent, so that the runtime tests can target the intended interface without enforcing a single reference solution. Marker placement was manually reviewed where the automatically chosen region did not correspond to a self-contained unit of behavior. The set of target files for a task is exactly the set of code files in its hollowed ROS~1 directory. 35 tasks have more than one target file, typically a header and its implementation, or a pair of nodes that must agree on an interface.\\

\subsection{Package Construction and Verification}
\label{sec:groundtruth}

A constructed task carries no reference ROS~2 implementation, no tests, and no environment in which either could run. All three are produced by the procedure described here, and none is assisted by the evaluated models.

For each task, we use Claude Sonnet 3.7 to assist in constructing the ROS~2 reference package, including the migrated target files, package configuration, auxiliary nodes, and an initial runtime test suite. The resulting artifacts are then built and executed inside the task-specific container. Upon failure, the error stage and logs are returned to the model for repair (up to three rounds). A task is frozen once its package builds and all tests pass, contributing three artifacts: the verified ROS~2 sources, the runtime test file, and the container definition. All 108 tasks were established successfully. The resulting artifacts are subsequently subjected to mutation-based test adequacy checks and manual audit against the original ROS~1 source.

The runtime tests exercise the migrated nodes rather than inspecting their source. A test starts the node under test as a subprocess through \texttt{ros2 run}, or imports the migrated module and instantiates the node in-process, and then interacts with it from a separate rclpy node created inside the test: publishing messages on the topics the node consumes, calling its services, reading or setting its parameters, and asserting on what is observed in return. Assertions are on concrete expected content rather than on the absence of exceptions, and each interaction is bounded by an explicit timeout with cleanup in a \texttt{finally} block.

\subsection{Execution Environment}

ROSTrans uses Docker environments to ensure that every task can be built and executed reproducibly. Each task has its own generated container definition, derived from a base image shared across the benchmark that provides the ROS~2 distribution and the dependencies needed to build and run the packages. The container copies the task's package into a workspace, builds it with \texttt{colcon build}, and, on success, runs the task's runtime tests with \texttt{pytest}. Build and execution are bounded by 900s and 200s respectively, and containers are run without network access so that concurrent evaluations cannot interfere with one another through middleware discovery traffic.

The container definition is generated from the same template for every task and is regenerated whenever the reference sources or tests change, so the environment in which a reference implementation was verified is also the environment in which candidates are later evaluated.

\subsection{Benchmark Profiling}
\label{sec:profiling}
 
\paragraph{Provenance}
The 108 tasks are derived from 70 distinct ROS1 repositories. The largest
contributors are \texttt{ros-planning/navigation} (8 tasks),
\texttt{ros/ros\_tutorials} (6) and \texttt{ros/ros\_comm} (4), followed by
\texttt{ROBOTIS-GIT/turtlebot3}, \texttt{moveit/moveit\_tutorials} and
\texttt{ros-industrial/reach\_ros} with 3 each; the remaining repositories
contribute one or two tasks, so no single upstream project dominates the
benchmark. The sources span navigation stacks, manipulation frameworks,
simulation bridges, sensor drivers, state-machine libraries and official
tutorial material. Provenance is recorded per task and released with the
benchmark; 15 tasks are adapted from tutorial material or built from minimal
ROS1 nodes and carry no upstream repository.
 
\paragraph{Scaffold shape}
Figure~\ref{fig:tasks} summarizes how much a model must supply. Tasks remove a median of 13 lines across 1.9 regions, growing with capability level. The number of files a model must return follows the same trend, from a mean of 1.11 at the interface level to 2.00 at the system level: 73 tasks have a single target file, 28 have two and 7 have three, and 28 of the 35 multi-file tasks are system-level, typically pairing a header with its implementation or two nodes that must agree on an interface. Because every target file is rewritten from scratch, a multi-file task additionally requires the returned files to remain mutually consistent.

\paragraph{Test suites}
The benchmark carries 976 runtime test functions, with a mean of 9.0, median of 6, and range of 1 to 38 per task. Suite size grows from 6.8 functions per interface-level task to 7.3 at the behavior level and 13.3 at the system level. The suites exercise the migrated nodes through exposed ROS~2 interfaces rather than by inspecting source text: 78.7\% of tasks launch the node under test as a separate process, 43.5\% subscribe to topics it publishes, 30.6\% publish to topics it consumes, 21.3\% read or write its parameters, and 17.6\% call its services. Smaller shares stand up a service the node must reach or drive an action interface, at 8.3\% each, while 7.4\% check transforms and 2.8\% exercise lifecycle transitions. Most tasks combine several mechanisms, and a test that only imports the module or checks that a process starts is not counted as exercising an interface.
 
\paragraph{Scope}
The tasks are middleware-level migrations. They exercise communication,
lifecycle, parameter and build-system behaviour, but no task moves a physical
robot or evaluates control quality, trajectory accuracy or perception
performance. Simulation appears only where a task's original package depends on
it, and then only to the extent needed to bring the node up.

% ==================================================================
\section{Evaluation Protocol}
\label{sec:eval}

Given a candidate's completion of a task, evaluation must place that completion into a runnable package, decide whether it works, and do so in a way that the candidate cannot influence. This section describes the injection procedure, the resulting outcome measure, and the evidence that the procedure itself is sound.

\subsection{Hybrid Injection}
\label{sec:injection}

A candidate returns its migrated files in a single response, marking each file with a \texttt{[FILENAME:~$f$]} header. Files are recovered by splitting the response on these headers and stripping any surrounding code fences.

Evaluation then assembles a temporary working copy. The reference package, its test directory, and its container definition are copied in full while only the candidate's versions of the target files are written over their counterparts. Because the target file set is determined by the task rather than by the response, a candidate cannot reach the tests or the build configuration even by naming them explicitly. This is the substantive difference from evaluations in which the generated artifact supplies or can influence its own tests: here, weakening the test is not an available strategy.

If a target file cannot be recovered from the response, or if the recovered name cannot be resolved to a file in the package, the task is scored as a failure; reference content is never substituted for a missing candidate file.

\subsection{Outcome and Metrics}

Outcomes are binary and are determined without any model in the loop. A task passes when the container builds, every test is collected, and none fails, errors, or is skipped; it fails otherwise. For each failure we record the stage at which it occurred, distinguishing build failures from test failures and from timeouts and unrecoverable output. 

Each task is attempted 5 times independently, and we report pass@$k$ using the unbiased estimator~\cite{chen2021evaluating}
\begin{equation}
    \text{pass@}k = \mathbb{E}_{\text{tasks}}\left[1 - \frac{\binom{n-c}{k}}{\binom{n}{k}}\right],
\end{equation}
where $c$ is the number of passing attempts for a task. We report 95$\% $ confidence intervals obtained by bootstrap resampling over tasks.
\subsection{Evaluation Validity}
\label{sec:validity}

\textbf{Evaluation Validity.} Three properties of the pipeline are enforced by construction. First, the generation prompt requires that tests exercise the translated files by name, through \texttt{ros2 run} or by importing the migrated module, and prohibits reimplementing target logic inside the test or replacing runtime interaction with textual inspection. Second, pass/fail is decided by a standalone program that parses the build result and the test summary, without models. Third, we verify these properties empirically through various checks.

\textbf{Test adequacy.} Because the reference tests are model-generated, we verify they constrain target behavior. Replacing the hollowed function with a no-op stub causes 94
of 108 tasks to fail immediately, confirming that their tests depend on the migrated implementation rather than on surrounding context, while the remaining 14 had tests strengthened until all 108 detect the stub. We further submit each reference implementation as a candidate through the full injection and scoring path (108
of 108 pass) and replace target files with non-functional content (103
of 108 fail).

\textbf{Manual audit.} A stratified sample of tasks was reviewed by hand against the original ROS~1 sources to check that the reference implementation preserves the intended behavior rather than merely satisfying its own tests.

\section{Experiment Setup}
\label{sec:experiment}

\paragraph{Evaluated Models.}
We evaluate 6 frontier LLMs covering proprietary, open-weight, general-purpose, and code-oriented model families from OpenRouter API. The reference implementations and runtime tests were produced
with Claude Sonnet 3.7, which is not among the evaluated
models, so no evaluated model contributed to the reference artefacts against
which it is scored.

\paragraph{Prompting Protocol.}
All models are evaluated in a zero-shot setting using a unified prompt format. Each prompt contains the scaffolded ROS~1 code, the task description, and the relevant migration context, and the model is instructed to return the completed contents of every file, each preceded by a \texttt{[FILENAME:]} header.

\paragraph{Generation and Evaluation.}
All models are queried through OpenRouter API with temperature set to 0.2. We run 5 independent generations per task under the same decoding setting, avoiding high-temperature sampling to reduce syntax hallucinations in ROS code. For each generation, we extract the returned files and inject them into the task's reference package as described in Section~\ref{sec:injection}. The resulting package is built and tested in the task's container. We then aggregate the results using pass@1, pass@3, and pass@5.

\section{Results}
\label{sec:results}

We evaluate six frontier LLMs on all 108 tasks with up to five stochastic generations per task, giving 3{,}240 evaluation runs of which 3{,}237 completed; three slots are missing because candidate generation produced no output, reported as absent rather than imputed. Every sample is scored by building the hybrid-injected
package and executing the task's runtime tests inside its own container, so a
sample passes only if the migrated code compiles and the nodes it defines
behave correctly at runtime.

\subsection{Overall Performance}

\begin{table}[ht]
\caption{pass@$k$ performance over 108 tasks (five samples/task).
Brackets give 95\% CIs from bootstrap resampling over the 108 tasks.}
\label{tab:main}
\centering\footnotesize
\setlength{\tabcolsep}{3pt}
\begin{tabular}{@{}lcrrrrr@{}}
\toprule
Model & pass@1 (\%) & pass@3 & pass@5 & \multicolumn{3}{c}{by level (\%)} \\
\cmidrule(l){5-7}
      & [95\% CI]   & (\%)   & (\%)   & Intf. & Behav. & Sys. \\
\midrule
Claude Opus 4.6 & 10.9 {\tiny[5.6,16.9]} & 12.5 & 13.0 &  8.2 & 11.7 & 13.7 \\
GPT-5.5         & 10.0 {\tiny[5.4,15.2]} & 14.6 & 16.7 &  9.5 & 12.4 &  8.6 \\
Gemini 3.1 Pro  &  7.2 {\tiny[3.5,11.5]} & 11.5 & 13.9 &  5.5 &  7.6 &  9.1 \\
Qwen 3.5 Plus   &  6.7 {\tiny[3.0,11.1]} &  9.3 & 10.2 &  6.4 &  3.4 &  9.7 \\
GLM-5 Turbo     &  6.5 {\tiny[2.6,10.9]} &  8.2 &  8.4 &  4.5 &  6.2 &  9.2 \\
DeepSeek V4 Pro &  5.7 {\tiny[2.2, 9.8]} &  8.6 &  9.4 &  4.6 &  3.5 &  9.1 \\
\bottomrule
\end{tabular}
\end{table}

Table~\ref{tab:main} reports pass@1 and pass@5 using the unbiased estimator of
Chen et al.~\cite{chen2021evaluating} over a fixed denominator of 108 tasks.
All six models solve fewer than one task in nine on a single attempt: pass@1
ranges from 5.7\% (DeepSeek V4 Pro) to 10.9\% (Claude Opus 4.6), and pass@5
from 8.4\% to 16.7\%. Aggregated over all 3{,}237 samples, 254 (7.8\%) pass.

Repeated sampling helps, but its shape differs across models. Two groups are
visible in Table~\ref{tab:main}. For Claude Opus 4.6 and GLM-5 Turbo, almost
all of the gain arrives by the third sample (10.9 to 12.5 to 13.0, and 6.5 to
8.2 to 8.4); additional samples reproduce the same failures rather than finding
new solutions. For GPT-5.5 and Gemini 3.1 Pro, the curve is still rising at
$k=5$ (10.0 to 14.6 to 16.7, and 7.2 to 11.5 to 13.9), indicating higher
sampling variance and a larger pool of tasks the model can solve intermittently.
Claude Opus 4.6 has the highest single-shot rate but the lowest relative gain
from sampling, so the ranking by pass@1 and by pass@5 is not the same. GPT-5.5 benefits most from repeated sampling (10.0\% to 16.7\%).

Grouping by capability level gives 6.4\% at the interface level, 7.5\% at the behavior level, and 9.9\% at the system level. These levels characterize architectural scope rather than a calibrated difficulty scale, so pass rates need not decrease monotonically with level. Figure~\ref{fig:category}
indicates why: the interface level contains the service-client and
parameter-server categories, which are among the weakest overall, while several
system-level tasks are C++ plugins whose tests check a narrower set of runtime
properties. We therefore read the level taxonomy as a description of
architectural scope rather than as a difficulty ordering. 

\begin{figure}[htbp]
  \centering
  \includegraphics[width=\linewidth]{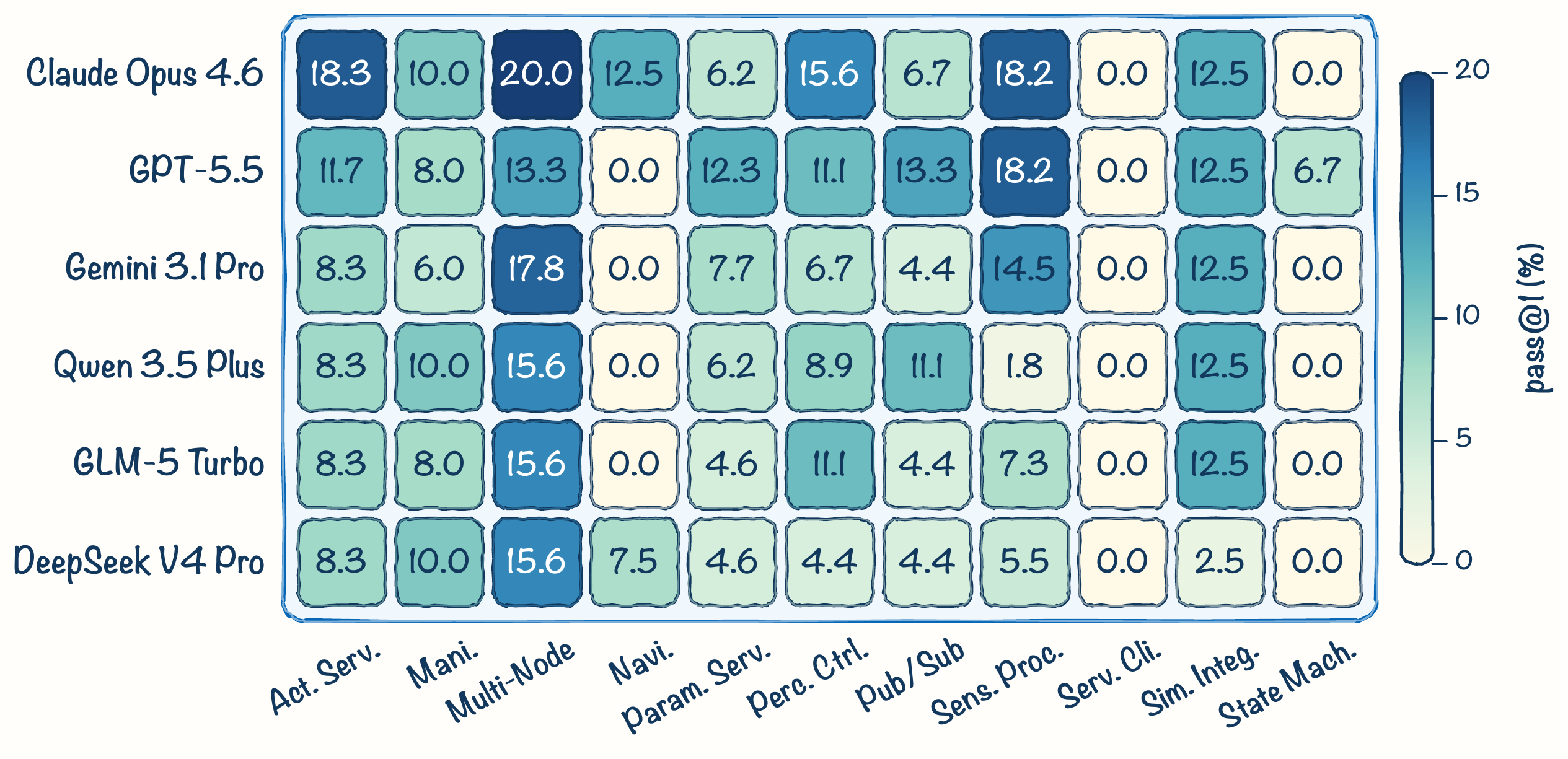}
  \caption{pass@1 across ROS capability categories, in percent. Columns are action server, manipulation, multi-node coordination, navigation, parameter server, perception and control, publisher/subscriber, sensor processing, service client, simulation integration, and state machine; rows are ordered by overall pass@1. Every model fails all service-client tasks, and the spread across categories is wider than the spread across models.}
  \label{fig:category_heatmap}
\end{figure}

\subsection{Where Solutions Fail}

\begin{figure}[htbp]
  \centering
  \includegraphics[width=\linewidth]{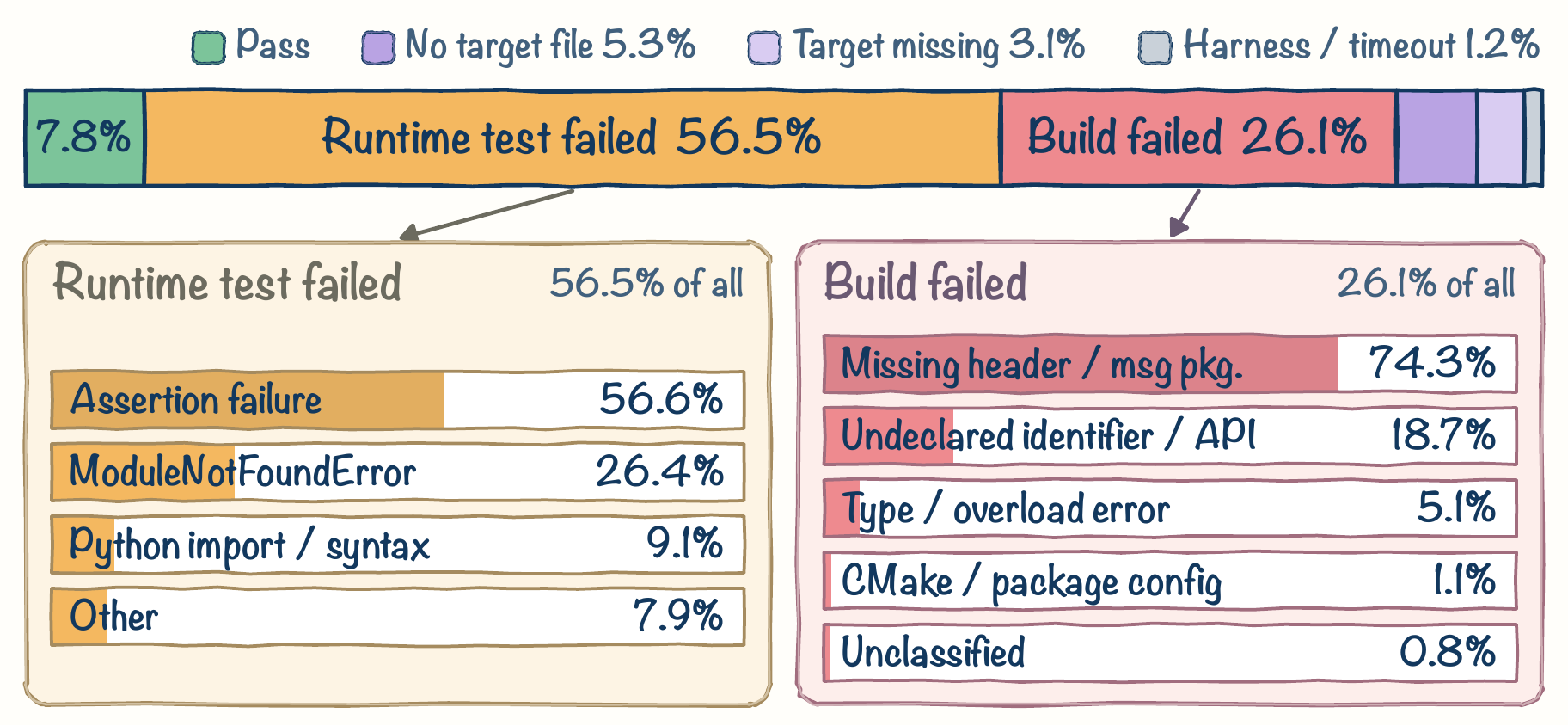}
  \caption{ Outcome of all 3,237 evaluated samples, with failures broken down
by stage and cause. More samples build successfully and then fail at
runtime than fail to build at all.}
  \label{fig:stages}
\end{figure}

Figure~\ref{fig:stages} breaks every sample down by its failure stage. The dominant failure mode is not compilation: 1{,}828 samples (56.5\%)
build successfully and fail runtime tests, against 844 (26.1\%) that
fail to build. More than twice as many samples reach successful compilation as fail during the build stage, yet most of these compilation-successful samples fail the runtime tests. An evaluation that stopped
at compilation would have scored those 1{,}828 samples as successes.

Classifying the 1{,}828 runtime failures by the exception that terminated the
test, 1{,}034 (56.6\%) are assertion failures: the package builds, the node
launches, and the test observes behaviour that contradicts the expected
interface or message contract. A further 483 (26.4\%) raise
\texttt{ModuleNotFoundError}, typically because a message package or a
placeholder package name is referenced in code but never declared in
\texttt{package.xml}, and 167 (9.1\%) fail on Python import or syntax errors.
Assertion failures are the category no static or compile-only check can reach:
the code is well-formed and the executable starts, and only exercising its
runtime interfaces reveals that the migration is wrong.

The build failures show the same weakness on the other side of the language
divide. Of the 844 samples that fail to build, 627 (74.3\%) terminate on a
missing header or message package, 158 (18.7\%) on an undeclared identifier or
a ROS2 API that does not exist, and 43 (5.1\%) on type or overload errors; only
7 (0.8\%) resist classification. Together with the Python
\texttt{ModuleNotFoundError} results above, this indicates that the models'
most frequent difficulty is not translating statements but reconstructing the
ROS2 package and dependency structure the translated statements require. We inspected the packages named in these failures. They are
overwhelmingly real ROS1 packages the original program
depended on (amcl, moveit, cv$\_$bridge, smach,
move$\_$base$\_$msgs), and ten samples still include \texttt{\#include <ros/ros.h>}
itself. A few recognise that a package was renamed under
ROS2 but guess the new name incorrectly. Models therefore
rarely invent APIs; their more frequent failure is not
recognising that a legacy dependency must be dropped or
replaced rather than carried forward.

A further 273 samples (8.4\%) never reach the build stage: 173 because no
target file could be extracted from the model's response, and 100 because the
response covered only part of the required target files. We count both as
failures and retain them in the denominator, since producing the requested
files in the requested form is part of the task. Harness errors and timeouts
account for the remaining 38 samples (1.2\%).

\subsection{Performance by Capability}

\begin{figure}[htbp]
  \centering
  \includegraphics[width=\linewidth]{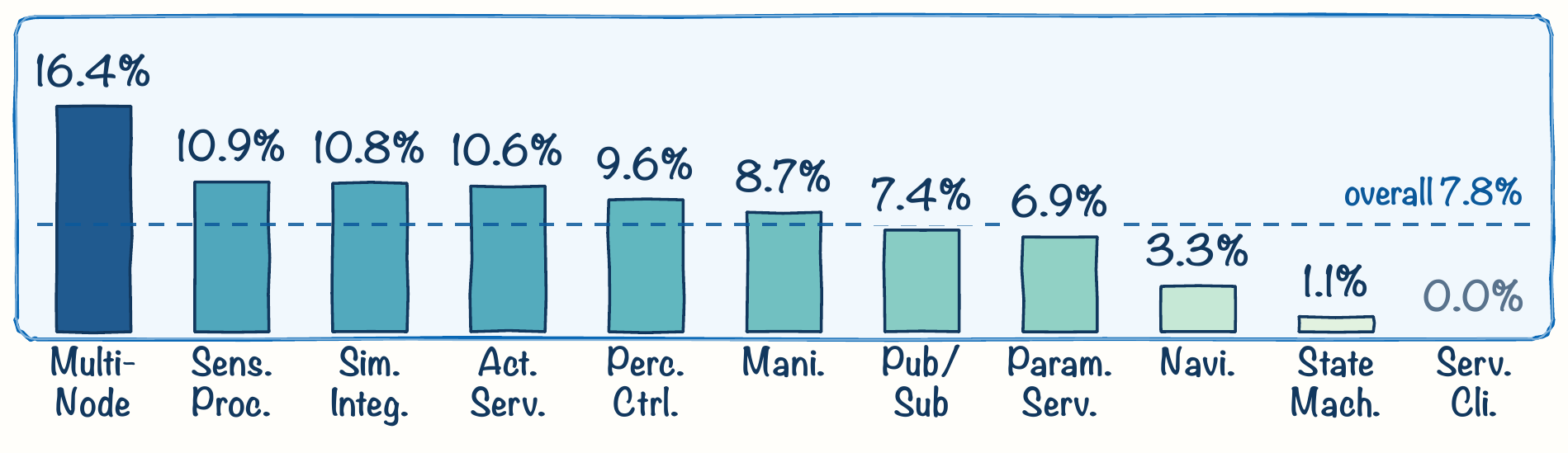}
  \caption{Pass rate by ROS capability category over all models and runs. Rates span 16.4$\%$ to zero, with service-client tasks passed by no sample.}
  \label{fig:category}
\end{figure}

Figure~\ref{fig:category} reports pass rates per ROS capability category
aggregated over all models and runs. The spread is wide. Multi-node
coordination is highest at 16.4\%, followed by sensor processing (10.9\%),
simulation integration (10.8\%) and action servers (10.6\%). At the other end,
navigation reaches 3.3\%, state machines 1.1\%, and no model passed any of the
300 service-client samples.

The service-client result deserves comment, since client--server migration is
often considered mechanical. Our tests require that the migrated node actually
advertise its service and answer a request, rather than merely instantiate a
client object. Of the 224 service-client samples that build but fail at
runtime, 78 (35\%) abort with a \texttt{ModuleNotFoundError} for a message
package or for a placeholder package name that was never declared, and a
further 9 fail because the node retains the ROS1
\texttt{\#!/usr/bin/env~python} shebang, which does not resolve under ROS2.
The remainder are assertion failures on the service response itself. To rule out that the category is unsolvable as specified rather than merely hard, two service-client tasks were solved by hand by an author, blind to the reference implementation, and passed on the first injection. The zero pass rate therefore reflects the candidates rather than an over-constrained test suite.

\subsection{Difficulty Distribution}

Per task rather than per sample, 85 of 108 tasks (78.7\%) were not solved by any model across five runs, with the remaining 23 accounting for all 254 passing samples. Current models succeed on a narrow band of
tasks and fail uniformly elsewhere, so the benchmark is far from saturated.

Because a near-uniform failure rate can also indicate a broken harness, we note
that the validity checks of Section~\ref{sec:validity} apply directly here: the
reference implementation passes in the identical container for every task, and
deliberately corrupting a target file causes the tests to fail. The zero-pass
tasks are therefore a property of the candidate solutions, not of the
evaluation infrastructure.

\subsection{Execution-Guided Repair}
\label{sec:iteration}
The zero-shot results measure single-shot migration, but practical workflows can provide execution feedback and let models iteratively repair their outputs. As a preliminary exploration, we evaluated Claude Opus 4.6 and GPT-5.5 on the 108 tasks with verified ground truth, allowing up to three rounds of execution-guided repair. The cumulative pass rate of Claude Opus 4.6 increased from 12/108 (11.1$\%$) to 32/108 (29.6$\%$), 44/108 (40.7$\%$) and 53/108 (49.1$\%$) over the three rounds, and GPT-5.5 followed a similar progression, from 12/108 (11.1$\%$) to 31/108 (28.7$\%$), 41/108 (38.0$\%$) and 51/108 (47.2$\%$). Qwen 3.5 Plus, a weaker baseline, gains almost all of its improvement in the first round, from 7/108 to 23/108, and saturates at 28/108 (25.9\%), so repair widens the gap between models rather than closing it. These gains of 38.0 and 36.1 percentage points suggest that execution feedback recovers a substantial fraction of the failures that remain unresolved in single-shot generation, motivating future work on systematic agentic migration, including the choice of feedback, repair strategies, and iteration budgets.

\section{Discussion}

\subsection{Shared Solutions and Run Consistency}
The six models together solve only 23 tasks, and 8 of them are solved by every model. Six tasks are solved by a single model, four by GPT-5.5 and two by Gemini 3.1 Pro; every task solved by Claude Opus 4.6, DeepSeek V4 Pro, GLM-5 Turbo or Qwen 3.5 Plus is also solved by at least one other model. An oracle that picks the best model per task would reach 21.3\%, only modestly above the best single model at 16.7\% pass@5, so current models fail on largely the same tasks rather than complementing one another. Run consistency explains the pass@$k$ shapes in Section~\ref{sec:results}: Claude Opus 4.6 solves 9 of its 14 solved tasks in all five runs and so gains little from sampling, whereas Gemini 3.1 Pro solves 6 of its 15 in only a single run.

\subsection{What Does and Does Not Predict Difficulty}
Task difficulty is not explained by scaffold size. Across tasks, the Spearman correlation between pass rate and the number of target files, TODO regions, hollowed lines, test functions or ROS~1 lines of code stays below 0.08 in magnitude. The ROS mechanism a task exercises is more telling. Tasks whose tests drive an action interface passed in 1 of 270 samples, lifecycle transitions in none, and transform checks in 2.9\%, while topic- and service-based tasks pass at 6.9\% to 12.6\%. Language shapes where failures surface rather than whether they occur: C++ tasks pass at 10.2\% and fail at the build stage in 46\% of samples, whereas Python tasks, which have no compilation step, pass at only 5.5\% and fail at runtime in 93\% of samples.

\subsection{Case Study: MP3 Inventory Service Semantic Failure}
\label{sec:case-study}

\begin{figure}[htbp]
  \centering
  \includegraphics[width=\linewidth]{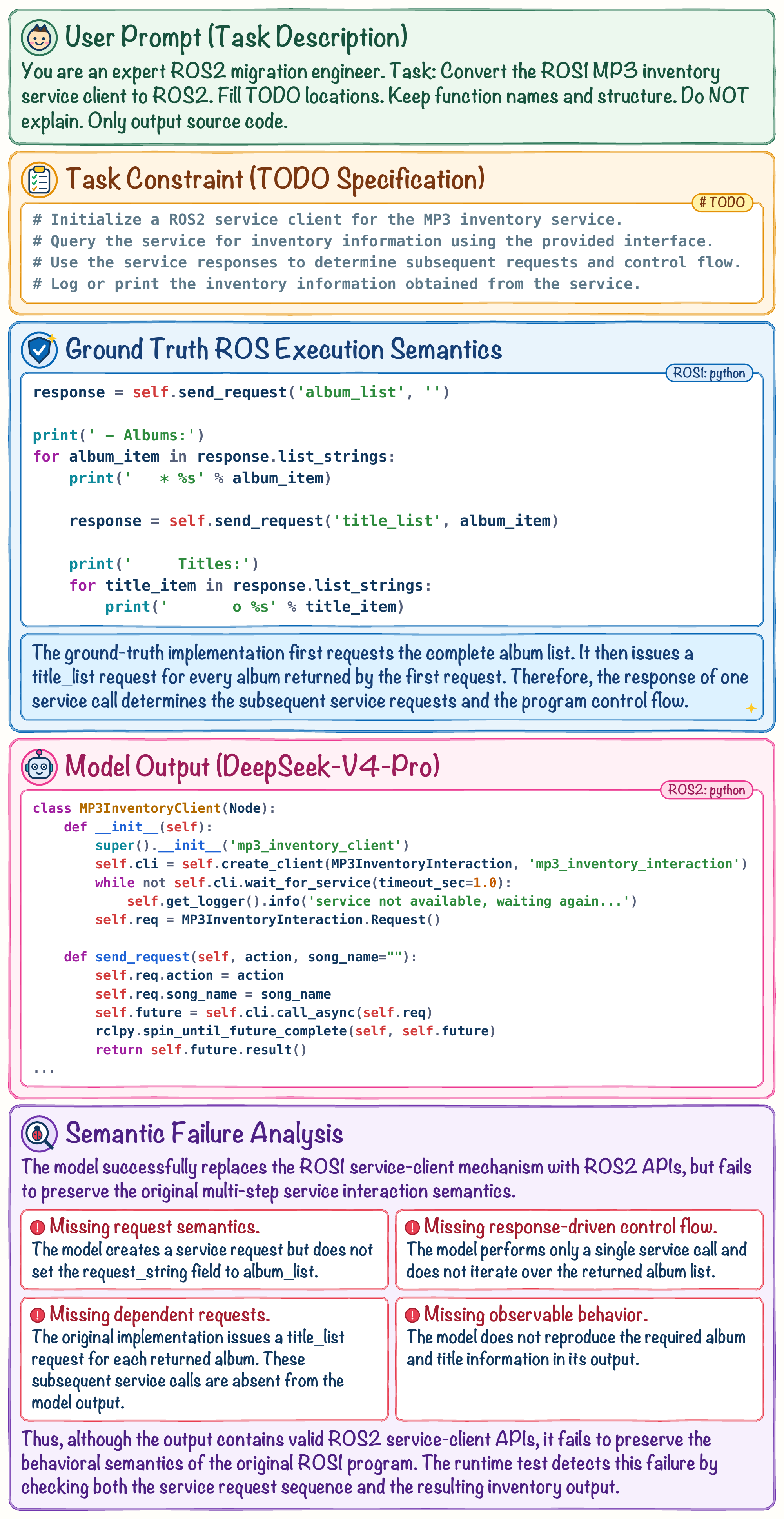}
  \caption {Case study of a service-client task. The model emits valid ROS2
service-client APIs but collapses the original two-stage interaction into
a single call, a semantic failure that only the runtime test detects.}
  \label{fig:case}
\end{figure}

To illustrate the gap between ROS API migration and semantic
preservation, we examine an MP3 inventory service migration case in Figure ~\ref{fig:case}.
The task requires migrating a ROS1 service client to ROS2 while completing a hollowed semantic core. The original program
performs a multi-step interaction: it retrieves the album list, then queries titles for each returned album. The model produces valid ROS2 service-client APIs but collapses the two-stage interaction into a single call, and the runtime test detects this by checking both the request sequence and the resulting output.

% \section{Discussion}
% \label{sec:discussion}

\section{Limitations and Future Work}
\label{sec:limitation}

Our execution-guided repair study is preliminary, covering two models and three repair rounds. A systematic evaluation across more models, feedback strategies, and iteration budgets remains important future work, and would better reflect industrial migration workflows, where software is developed, tested, and revised over multiple cycles. The fixed build configuration is part of the task definition rather than an oversight: it keeps the build configuration outside the candidate's reach so that tests cannot be weakened. Section~\ref{sec:iteration} bounds its impact, since 49.1$\%$ of tasks are solved under the identical configuration. Finally, the current tasks focus on middleware-level correctness; future extensions could incorporate control, perception, and physical-robot evaluation.

\section{Conclusion}
\label{sec:conclusion}
We introduce ROSTrans, an execution-grounded benchmark for ROS1-to-ROS2 code migration, where correctness depends on communication patterns, middleware behavior, and distributed node interaction rather than surface-level API replacement. ROSTrans organizes 108 tasks into interface, behavior, and system level migration and evaluates each task by building the migrated package and exercising its nodes at runtime inside a container. Experiments on six frontier LLMs show that current models solve fewer than one task in nine on a single attempt and gain only modestly from repeated sampling. Most failing migrations build successfully but fail at runtime, showing that compilation alone substantially overestimates migration correctness, and missing or undeclared package dependencies remain a persistent source of failure. These results highlight the gap between API-level translation and system-level behavioral correctness.

\bibliographystyle{IEEEtran}
\bibliography{refs}

\end{document}